\documentclass[aps,pra,twocolumn,showpacs,superscriptaddress,longbibliography]{revtex4-2}
\usepackage{graphicx} 
\usepackage{amsmath}
\usepackage{multirow}
\usepackage{graphicx,epstopdf}
\usepackage{booktabs}
\usepackage{gensymb}
\usepackage[dvipsnames]{xcolor}
\usepackage{footnote}
\usepackage{multibib}

\newcommand*{\hatH}{\hat{\mathcal{H}}}

\newcommand{\be}{\begin{equation}}
	\newcommand{\ee}{\end{equation}}
\newcommand{\bea}{\begin{eqnarray}}
	\newcommand{\eea}{\end{eqnarray}}
\newcommand{\bse}{\begin{subequations}}
	\newcommand{\ese}{\end{subequations}}

\graphicspath{{../}}
\usepackage{color}
\usepackage[colorlinks,bookmarks=false,citecolor=darkblue,linkcolor=red,urlcolor=blue]{hyperref}

\definecolor{darkred}{rgb}{0.7,0.0,0.0}

\definecolor{darkblue}{rgb}{0,0.02,0.45}

\definecolor{darkgreen}{rgb}{0.02,0.45,0.0}

\definecolor{violet}{rgb}{0.8,0.2,0.6}

\usepackage{mathtools}

\DeclarePairedDelimiter\ket{\lvert}{\rangle}
\DeclarePairedDelimiterX\braket[2]{\langle}{\rangle}{#1\,\delimsize\vert\,\mathopen{}#2}

\begin{document}

\title{Disordered ground state in the spin-orbit coupled $J_{\rm eff}= 1/2$ cobalt-based metal-organic framework magnet with orthogonal spin dimers}	
\author{Sebin J. Sebastian}
\affiliation{School of Physics, Indian Institute of Science Education and Research Thiruvananthapuram-695551, India}
\author{S. Mohanty}
\affiliation{School of Physics, Indian Institute of Science Education and Research Thiruvananthapuram-695551, India}
\author{A. Nath}
\affiliation{School of Chemistry, Indian Institute of Science Education and Research Thiruvananthapuram-695551, India}
\author{M. P. Saravanan}
\affiliation{UGC-DAE Consortium for Scientific Research, University Campus, Khandwa Road, Indore 452001, India}
\author{S. Mandal}
\affiliation{School of Chemistry, Indian Institute of Science Education and Research Thiruvananthapuram-695551, India}
\author{A. A. Tsirlin}
\affiliation{Felix Bloch Institute for Solid-State Physics, Leipzig University, 04103 Leipzig, Germany}
\author{R. Nath}
\email{rnath@iisertvm.ac.in}
\affiliation{School of Physics, Indian Institute of Science Education and Research Thiruvananthapuram-695551, India}
\date{\today}
\begin{abstract}
We present the magnetic properties of a strongly spin-orbit coupled quantum dimer magnet based on Co$^{2+}$. The metal-organic framework compound Co$_2$(BDC)$_2$(DPTTZ)$_2$$\cdot$DMF features Co$^{2+}$ dimers arranged nearly orthogonal to each other, similar to the Shastry-Sutherland lattice. Our assessment based on the magnetization and heat capacity experiments reveals that the magnetic properties at low temperatures can be described by an effective $J_{\rm eff} = 1/2$ Kramers doublet and the ground state is a singlet with a tiny spin gap. Although the magnetic susceptibility could be analyzed in terms of the interacting dimer model with an isotropic intra-dimer coupling $J_0/k_{\rm B} \simeq 7.6$~K, this model fails to reproduce the shape of magnetization isotherm and heat capacity data. A model of isolated spin dimers with the anisotropic exchange couplings $J_{xy} \simeq 3.5$~K and $J_{z} \simeq 11$~K provides an adequate description to the magnetic susceptibility, magnetization isotherm, and heat capacity data at low temperatures. Interestingly, no field-induced quantum phase phase is detected down to 100~mK around the critical field of gap closing, suggesting the absence of Bose-Einstein condensation of triplons and establishing isolated dimers with a negligible interdimer coupling. 
\end{abstract}
\maketitle

\section{Introduction}
Quantum phase transition (QPT) in a spin system is a transition between two phases at absolute zero temperature driven by quantum fluctuations~\cite{Sachdev33,*Vojta2069}. Such a phase can be accessed upon variation of an external control parameter that has direct bearing on the system properties.
To study QPT, antiferromagnetic (AFM) spin-$1/2$ dimers are well suited systems, which are characterized by a singlet ($\ket{S,S_z}=\ket{0,0}$) ground state with entangled spins and an excitation gap in the energy spectrum~\cite{Zapf563}. Application of external magnetic field leads to a closing of the excitation gap and breaking of SU(2) symmetry that in turn stabilizes the magnetic long-range order (LRO).
This non-trivial field-induced LRO can be described well as the Bose-Einstein condensation (BEC) of spin-1 triplons~\cite{Rice760,Zapf563,Giamarchi198}. Experimentally, BEC physics is demonstrated in compounds featuring coupled spin dimers, even-leg ladders, and bond-alternating spin chains~\cite{Nikuni5868,Jaime087203,Thielemann020408, Garlea167202,Mukharjee144433,Mukharjee224403}. Another fascinating field-induced quantum effect is the appearance of magnetization plateaus, typically observed in spin systems with orthogonal dimers, such as in the Shastry-Sutherland lattice SrCu$_2$(BO$_3$)$_2$~\cite{Kodama395,Kageyama3168}.

In the past, oxide materials with several transition-metal and few rare-earth ions (e.g. Yb$^{3+}$) are studied in the context of field-induced transitions~\cite{Laflorencie060602,Suh054413,Freitas184426,Lancaster207201,Zapf077204,Garlea167202,Hester027201}. Among these systems, Co$^{2+}$-based compounds are quite attractive as they evince peculiar features, such as the $J_{\rm eff}= 1/2$ Kramers-doublet ground state, similar to Yb$^{3+}$ ($4f$) and Ce$^{3+}$ ($4f$) magnets~\cite{Hester027201,Somesh064421,Mohanty134408}, single-ion anisotropy~\cite{Abragam173} etc.
Furthermore, strong spin-orbit coupling (SOC) and crystal electric field (CEF) in Co$^{2+}$ systems introduce bond-dependent anisotropy of the exchange couplings, thus offering a promising playground to realize exotic phases, including field-induced quantum spin liquid (QSL) in Kitaev materials~\cite{Lin5559,Liu054420}, BEC of triplons~\cite{Sheng2211193119}, and Berezinskii-Kosterlitz-Thoueless (BKT) transition~\cite{Gao89}.
Recently, BEC of triplons is observed in a $4f$-based dimer magnet Yb$_2$Si$_2$O$_7$ but with an unusually asymmetric BEC dome~\cite{Hester027201}. This asymmetry is sometimes attributed to the anisotropy in the microscopic exchanges in the Hamiltonian~\cite{Flynn067201,Feng205150}. Similar physics is also envisaged for Co$^{2+}$-based compounds because of the exchange anisotropy~\cite{Zhang10381}.
In this regard, Co$^{2+}$-based metal-organic compounds are ideal contenders and provide a convenient testing ground for field-induced studies due to reduced exchange couplings compared to their inorganic counterparts. This leads to lower critical fields, enabling the experimental techniques to have access to the complete phase diagram using continuous-field laboratory magnets~\cite{Freitas184426,Coak12671,Lancaster207201}.

\begin{figure}[h]
	\includegraphics[width= \linewidth]{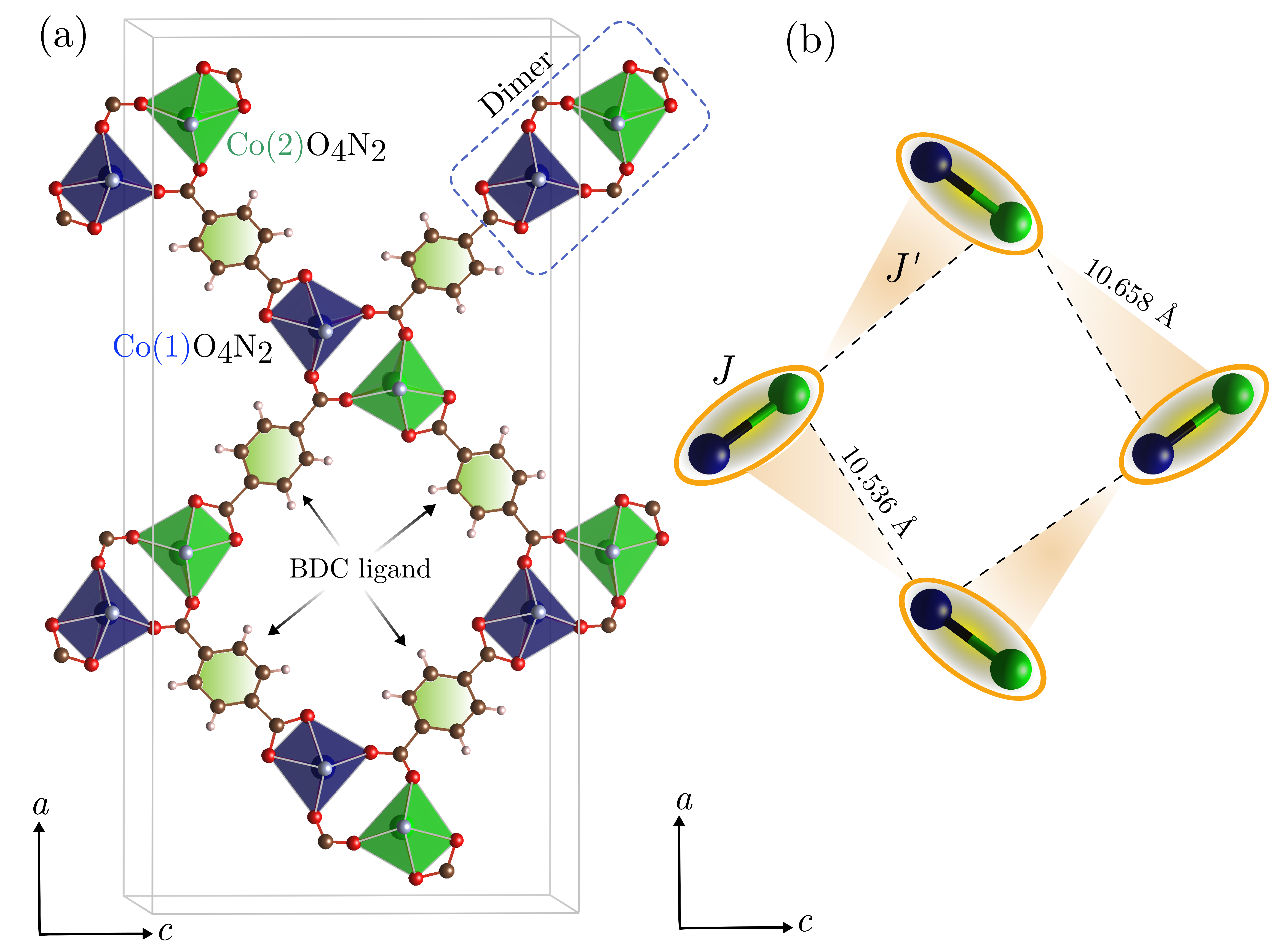}
	\caption{\label{Fig1} (a) Structure of Co-MOF in one unit cell showing the formation of spin dimers by the CoO$_4$N$_2$ units and the interdimer connectivity. (b) Schematic diagram showing the orthogonal arrangement of dimers and the possible exchange couplings.}
\end{figure}

In this paper, we report the structural and magnetic properties of a Co$^{2+}$-based metal-organic framework Co$_2$(BDC)$_2$(DPTTZ)$_2$$\cdot$DMF (abbreviated as Co-MOF), investigated via magnetization and heat capacity measurements down to 100~mK, followed by exact diagonalization calculations. Co-MOF crystallizes in an orthorhombic space group $Iba2$ (No.~45) with the lattice parameters $a=34.7269(1)$~Å, $b=17.1760(3)$~Å, $c=17.0489(5)$~Å, and $\alpha=\beta=\gamma=90^\circ$~\cite{Zhai7173,Nath202308034}. It contains two crystallographically in-equivalent cobalt atoms [Co(1) and Co(2)].
Each Co(II) center is coordinated with O and N atoms in a distorted heteroleptic octahedron (CoO$_4$N$_2$) geometry with Co-O and Co-N distances in the ranges of $1.988(4)-2.232(4)$ Å and $2.134(3)-2.160(3)$ Å, respectively. Two in-equivalent cobalt octahedra are connected via two C atoms and
form a Co(1) -- Co(2) dimer. These dimer units are further connected through phenyl ring of benzene dicarboxylate (BDC) ligands to form a corrugated two-dimensional layer [Fig.~\ref{Fig1}(a)]. Each layer is pillared by DPTTZ ($N, N'$-di(4-pyridyl)thiazolo-[5,4-d] thiazole) units to form an overall three-dimensional framework. Interestingly, these Co-Co dimers are arranged in an orthogonal fashion, reminiscent of the famous Shastry-Sutherland lattice [see Fig.~\ref{Fig1}(b)]. The possible intradimer interaction $J$ ($d_{\rm Co-Co}\simeq4.04$~Å) and interdimer interaction $J'$ ($d_{\rm Co-Co}^{\rm avg}\simeq10.597$~Å) are also depicted in Fig.~\ref{Fig1}(b).

\section{Methods}
The single-crystal synthesis of Co-MOF involves two steps. In the first step, the precursor DPTTZ was prepared by mixing 200~mg of dithiooxamide and 0.4~mL of 4-pyridinecarboxaldehyde in 10~mL of high-performance liquid chromatography $N,N'$-dimethylformamide (DMF), and refluxed at 153$^\circ$C for 5 h~\cite{Nath2227}. The yellow-colored needle-shaped crystals of DPTTZ were formed after slow cooling. The crystals were washed in water, dried under vacuum, and stored under ambient conditions. In the second step, 0.027~mmol (7.85~mg) Co(NO$_3$)$_2$.6H$_2$O, 0.03~mmol (4.98~mg) terephthalic acid, and 0.025~mmol (7.4~mg) DPTTZ were mixed in 2:0.5 DMF : H$_2$O ratio and sonicated for 1~h. The resulting solution was transferred to a teflon lined vessel and kept in an autoclave at 100$^\circ$C for two days. Co-MOF crystals were obtained after slow cooling of the autoclave. The crystals were washed several times with DMF and dried under vacuum. The crystals were maintained under ambient conditions prior to further experiments.

\begin{figure}[h]
	\includegraphics[width= \linewidth]{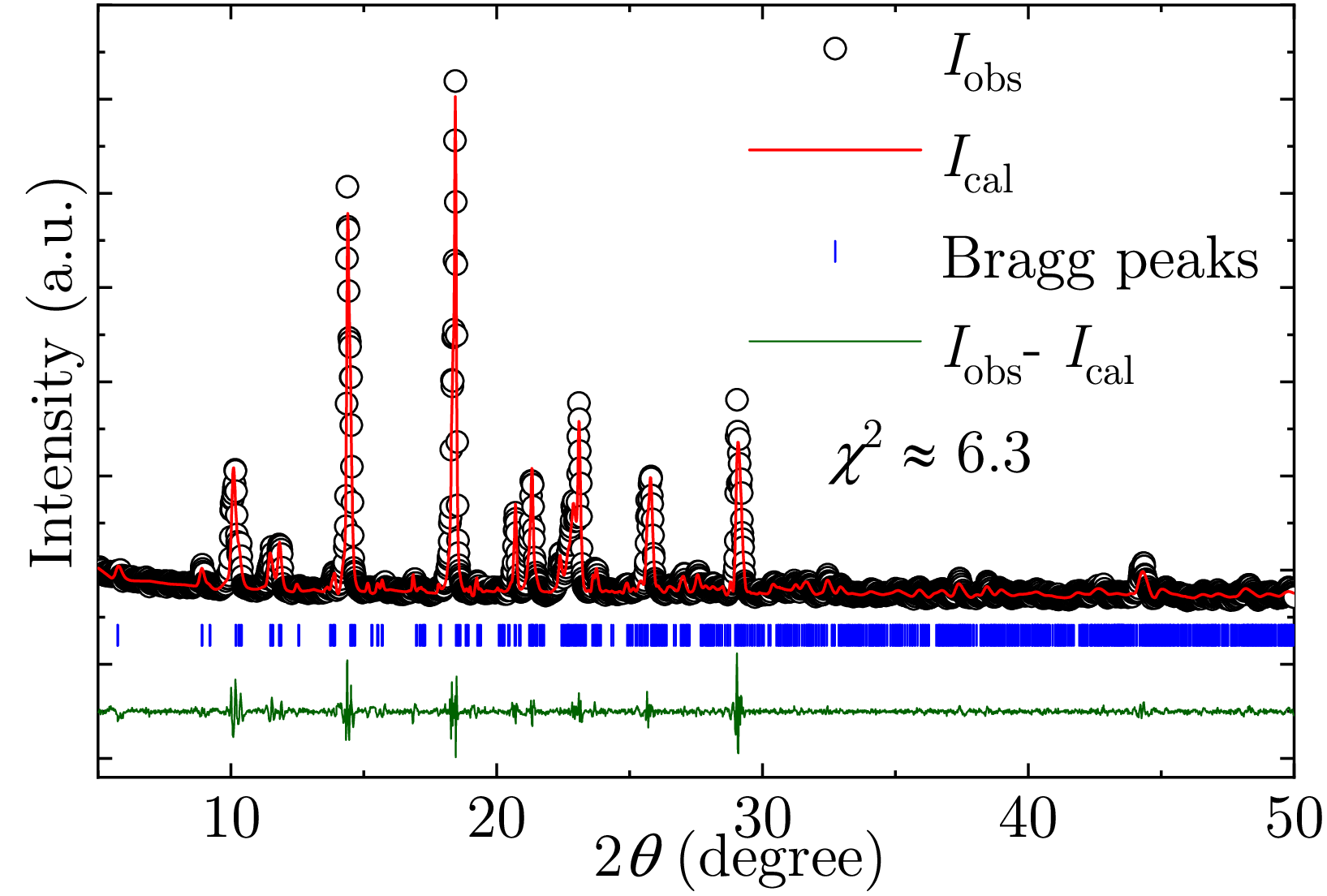}
	\caption{\label{Fig2} Powder XRD pattern of Co-MOF measured at room temperature. The open circles represent the experimental data and the solid red line represents the Le-Bail fit. The Bragg positions are indicated by blue vertical bars and the bottom solid green line indicates the difference between the experimental and calculated intensities.}
\end{figure}
A large number of small single crystals were crushed into fine powder and the phase purity of the product was confirmed via powder x-ray diffraction (XRD) at room temperature using a PANalytical x-ray diffractometer (Cu\textit{K$_{\alpha}$} radiation, $\lambda_{\rm avg}\simeq 1.5418$~\AA). Le-Bail analysis of the powder XRD pattern was performed using the \texttt{MAG2POL} software package~\cite{Qureshi175}. Figure~\ref{Fig2} shows the powder XRD data along with the fit at $T = 300$~K. The structural parameters given in Ref.~\cite{Zhai7173} were used as initial parameters. The obtained lattice parameters $a=34.9769(1)$~Å, $b=17.1990(1)$~Å, $c=17.1076(1)$~Å, and $\alpha=\beta=\gamma=90^\circ$, are in close agreement with the previous report~\cite{Zhai7173}.

DC magnetization ($M$) of the powder sample was measured using a superconducting quantum interference device (SQUID) magnetometer (MPMS-3, Quantum Design). The measurements were performed in the temperature range of 1.8~K~$\leq T \leq 380$~K and in the magnetic field range of 0~T~$\leq H \leq 7$~T. The SQUID enabled us to measure the isothermal magnetization at $T = 0.4$~K with the help of a $^3$He attachment. Heat capacity ($C_{\rm p}$) was measured on a small piece of sintered pellet using the thermal relaxation technique over a large temperature range in two physical property measurement systems (PPMS, Quantum Design). For $T > 2$~K, measurements were done in a 9~T PPMS while for 0.1~K $\leq T \leq  4$~K we have used a dilution insert in a 14~T PPMS.

The magnetization and heat capacity of the anisotropic spin dimer are simulated via exact diagonalization using the \texttt{fulldiag} utility of the \texttt{ALPS} package~\cite{Bauer5001}, where the number of sites is taken to be $Z = 2$.

\section{Results}
\subsection{Magnetization}
\label{sec:magnetization}
\begin{figure}[h]
\includegraphics[width= \linewidth]{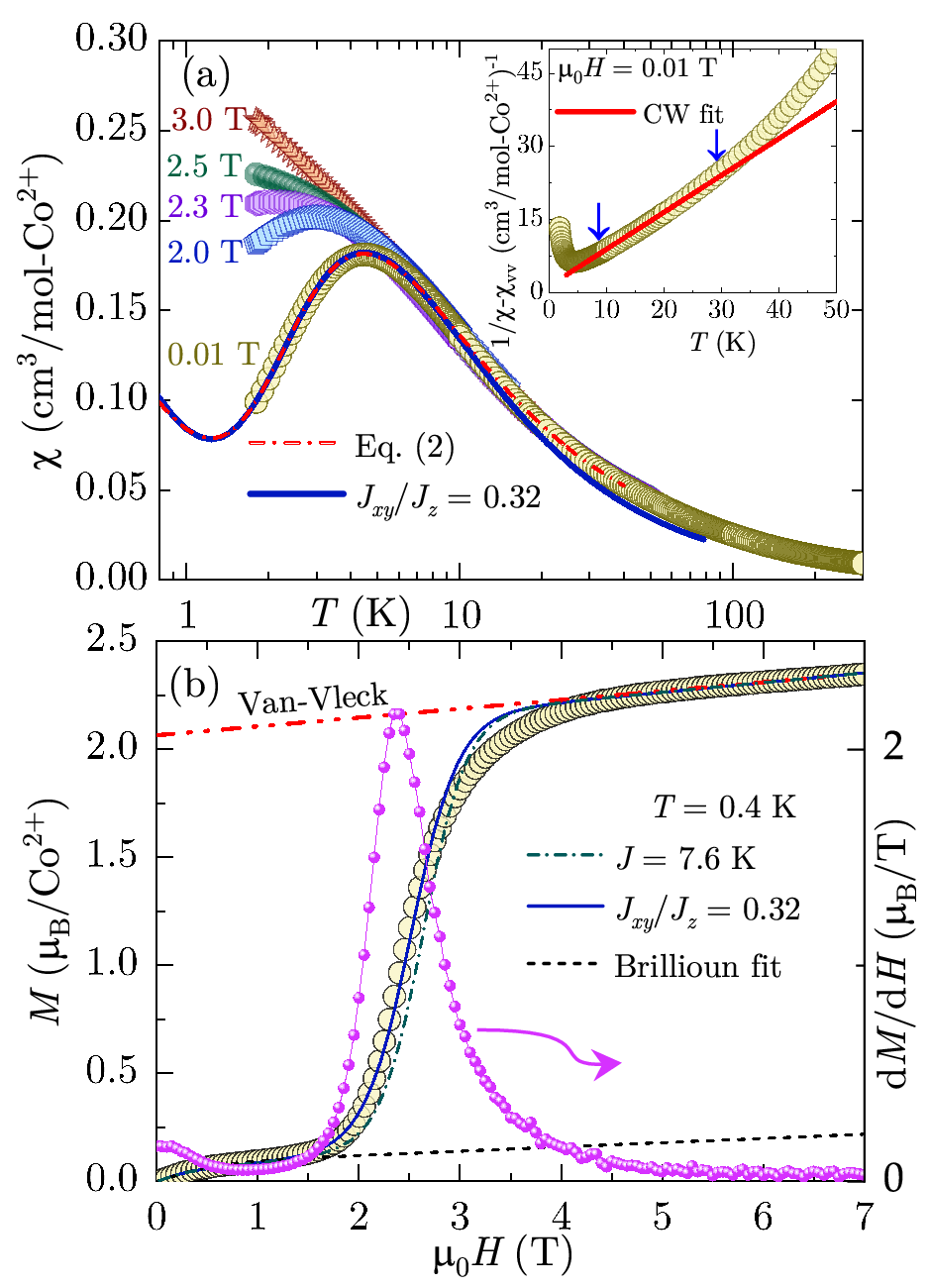}
\caption{\label{Fig3}(a) $\chi(T)$ of Co-MOF measured in different applied fields. The dash-dotted line represents the fit of the $\mu_0 H = 0.01$~T data by the interacting dimer model [Eq.~\eqref{eq2}] while the solid line is the simulation of isolated spin dimers with anisotropic interactions ($J_{z} = 11$~K and $J_{xy} = 3.52$~K). Inset: CW fit to the low-$T$ $1/\chi$ data (after subtracting the van-Vleck contribution). (b) $M$ vs $H$ and $dM/dH$ vs $H$ on the left and right $y$-axes, respectively, measured at $T = 0.4$~K. The horizontal dash-dotted line marks the van-Vleck contribution. The dashed line shows the Brillouin fit [Eq.~\eqref{BF}] to magnetization in the gaped regime. The dash-dotted and solid lines are the simulations for isolated spin dimers with isotropic and anisotropic interactions, respectively.}
\end{figure}

Temperature-dependent magnetic susceptibility $\chi~(\equiv M/H)$ measured in different magnetic fields is shown in Fig.~\ref{Fig3}(a).
At the lowest measured field ($\mu_0 H = 0.01$~T), $\chi(T)$ passes through a broad maximum at $T^{\rm max}_\chi \simeq 3$~K, which is the signature of an antiferromagnetic (AFM) short-range order, typical for a low-dimensional spin system. With increasing field, the position of the broad maximum shifts towards low temperature, as expected. Below 3~K, $\chi(T)$ exhibits a rapid decrease, suggesting a singlet ground state or the opening of a spin gap, which necessitates additional investigation~\cite{Arjun014421,Ahmed224423}. No anomaly associated with the magnetic LRO is observed down to 1.8~K. 

To extract key magnetic parameters, we conducted the $\chi(T)$ analysis employing the modified Curie-Weiss (CW) law:
\begin{equation}
\chi(T) = \chi_0 + \frac {C}{T-\theta_{\rm CW}},
\label{eq1}
\end{equation}
where $\chi_{0}$ is the temperature-independent contribution consisting of the core diamagnetic susceptibility ($\chi_{\rm core}$) and van Vleck paramagnetism ($\chi_{\rm VV}$) of the open shell of the Co$^{2+}$ ions present in the sample. The second term in Eq.~\eqref{eq1} represents the CW law with the Curie constant $C=N_{\rm A} \mu_{\rm eff}^2/3k_{\rm B}$, where $N_{\rm A}$ is the Avogadro number, $k_{\rm B}$ is the Boltzmann constant, $\mu_{\rm eff}=g\sqrt{S(S+1)}\mu_{\rm B}$ is the effective magnetic moment, $g$ is the Land\'e $g$-factor, $\mu_{\rm B}$ is the Bohr magneton, $S$ is the spin quantum number, and $\theta_{\rm CW}$ is the characteristic CW temperature. Above 150~K, $\chi(T)$ for $\mu_{0}H=0.01$~T was fitted by Eq.~(\ref{eq1}) that yields $\chi_{0}\simeq 7.3\times10^{-4}$~cm$^3$/mol, the high-$T$ (HT) effective moment $\mu_{\rm eff}^{\rm HT}\simeq 4.53~\mu_{\rm B}$/Co$^{2+}$, and the high-$T$ CW temperature $\theta_{\rm CW}^{\rm HT} \simeq-4~$K. The value of $\mu_{\rm eff}^{\rm HT}$ corresponds to $S = 3/2$ with $g\simeq2.33$ for the $3d^7$ configuration. A slightly larger $\mu_{\rm eff}^{\rm HT}$ value compared with the theoretical spin-only value ($\sim 3.87~\mu_{B}$/Co$^{2+}$ for $S=3/2$ with $g=2.0$) indicates anisotropy in $g$-factor stemming from the pseudo-trigonal crystal field as well as additional orbital contribution, as observed in several Co$^{2+}$-based systems~\cite{Eichhofer1962,Iakovleva094413}.

At temperatures below 40~K, $1/\chi$ exhibits a subtle slope change, deviating from the high-$T$ CW behavior. Spin-orbit coupling splits the Co$^{2+}$ multiplet into several Kramers doublets, and the physics at low temperatures is fully determined by the lowest Kramers doublet that can be treated as pseudospin-$\frac12$. This type of crossover from the high-$T$ $S=3/2$ to the low-$T$ $J_{\rm eff}=1/2$ behavior is a common character of Co$^{2+}$-based compounds~\cite{Lines546,Susuki267201,Lal014429}. To understand the low-temperature $\chi(T)$ data precisely, one requires a proper estimation of $\chi_{\rm VV}$ that arises from excitations between the ground-state Kramers doublet and excited states. After subtracting $\chi_{\rm VV}$ (obtained from the magnetization isotherm at $T=0.4$~K) from $\chi(T)$, $1/(\chi-\chi_{\rm VV}$) shows a clear slope change and a linear regime below 40~K [inset of Fig.~\ref{Fig3}(a)]. A CW fit [Eq.~\eqref{eq1}] in the low-$T$ (LT) regime (8~K~-~30~K) returns $\mu_{\rm eff}^{\rm LT}\simeq 3.40~\mu_{\rm B}$/Co$^{2+}$ and $\theta_{\rm CW}^{\rm LT} \simeq-1.9$~K. The $\mu_{\rm eff}^{\rm LT}$ value corresponds to a $J_{\rm eff}= 1/2$ moment with an average $g \simeq 3.93$. Such a large value of $g$ compared to the free electron value $g=2$ reflects the effect of spin-orbit coupling~\cite{Li4216,Ranjith115804,Lal014429,Susuki267201}. The negative value of $\theta_{\rm CW}^{\rm LT}$ confirms the dominant AFM interactions at low temperatures.

From the crystal structure, while considering the spin-lattice of Co atoms, we anticipate the formation of magnetic dimers of Co$^{2+}$ ions in an orthogonal fashion, reminiscent of Shastry-Sutherland lattice. As a first step, to determine the potential exchange couplings, we fitted the $\chi(T)$ data using the following expression: 
\begin{equation}
\chi(T)=\chi_0 + \frac{C_{\rm imp}}{T} + \chi_{\rm d}(T).
\label{eq2}
\end{equation}
Here, the second term is the Curie law and $\chi_{\rm d}(T)$ is the expression for susceptibility of AFM spin-1/2 coupled dimer model which has the form~\cite{Arjun174421}
\begin{equation}
\chi_{\rm d} (T)= \frac{N_{\rm A}g^2\mu_{\rm B}^2}{k_{\rm B}T [{3+\exp({J/k_{\rm B}T})+zJ{^\prime}/k_{\rm B}T}]}.
\label{eq3}
\end{equation}
Here, $J/k_{\rm B}$ is the intra-dimer exchange coupling, $J^\prime/k_{\rm B}$ represents the inter-dimer exchange coupling, and $z$ is the number of nearest-neighbor spins. The fit using Eq.~\eqref{eq2} in the low temperature regime (1.80~K~$\leq~T\leq$~40~K) [see Fig.~\ref{Fig3}(a)] returns $\chi_0 \simeq 0.0132$~cm$^3$/mol, $C_{\rm imp}\simeq 0.0678$~cm$^3$-K/mol, $g \simeq 4.15$, $z=2$, $J/k_{\rm B} \simeq 7.6$~K, and $J{^\prime}/k_{\rm B} \simeq 0.07$~K. The obtained value of $C_{\rm imp}$ corresponds to $\sim 4.1$\% of spin-1/2 impurities. The value of $J/k_{\rm B}\simeq 7.6$~K with $g \simeq 4.15$ sets the critical field of gap closing $H_{\rm C1} = \Delta_0/g\mu_{\rm B} \simeq 2.7$~T where $\Delta_0\simeq J/k_B$ is the spin gap between the singlet and triplet states of the dimer~\cite{Mukharjee144433}.

To track the magnetization response with respect to the applied field, we measured an isothermal magnetization [$M(H)$] at $T=0.4$~K up to 7~T. For a gapped spin system, magnetization ($M$) typically remains zero below $H_{\rm C1}$, while above this field it is expected to increase up to the saturation field $H_{\rm C2}$ that marks the onset of the fully polarized state~\cite{Tsirlin144412,Nohadani024440}. As shown in Fig.~\ref{Fig3}(b), in the low-field regime, $M$ exhibits a wide plateau up to $H \sim 2$~T, clearly suggesting a singlet ground state and the critical field required to overcome the spin gap is about $H_{\rm C1} \simeq 2$~T. This plateau with non-zero magnetization can be attributed to the paramagnetic impurities/defects expected in a powder sample, which saturate above 1~T. Above $H_{\rm C1} \simeq 2 $~T, $M$ increases sharply and saturates in high fields. It is also observed that $M$ increases weakly with increasing field beyond the saturation field $H_{\rm C2}$. This weak increase in high fields is primarily due to the van Vleck contribution ($\chi_{\rm VV}$), expected for Co$^{2+}$ systems in an octahedral environment~\cite{Shiba2326}. The slope and $y$-intercept of a linear fit to the data above $\sim 4$~T yields $\chi_{\rm VV} \simeq 2.46\times10^{-2}$~cm$^3$/mol and the saturation magnetization of $M_{\rm S}\simeq 2.05~\mu_{\rm B}$/f.u., respectively. This value of $M_{\rm S}$ is in good agreement with $M_{sat} = gJ_{\rm eff} \mu_{\rm B} \simeq 2.0~\mu_{\rm B}$ expected for $J_{\rm eff} = 1/2$ with powder averaged $g\simeq 4.0$. Thus, the magnetization isotherm analysis also validates the $J_{\rm eff} = 1/2$ ground state of Co$^{2+}$.

In order to quantitatively estimate the impurity contribution and to visualize zero magnetization below $H_{\rm C1}$, we fitted the data below 2~T [see Fig.~\ref{Fig3}(b)] by the following equation~\cite{Mukharjee144433}
\begin{equation}
M=\chi H + N_{\rm A}\mu_{\rm B} f_{\rm imp}S_{\rm imp}g_{\rm imp}\times B(x),
\label{BF}
\end{equation}
where $\chi$ is the intrinsic susceptibility of the sample, $f_{\rm imp}$ is the molar fraction of the impurities, $g_{\rm imp}$ is the impurity $g$-factor, $S_{\rm imp}$ represents the impurity spin, $B(x)$ is the Brillouin function taken for the spin value $S_{\rm imp}(x)$, and the argument ($x$) can be defined by $x= g_{\rm imp}\mu_{\rm B}S_{\rm imp}H/[k_{\rm B} (T+\theta_{\rm imp})]$. The fit yields the following parameters: $f_{\rm imp} \simeq 0.039$, $S_{\rm imp} = 0.5$, $g_{\rm imp} \simeq 4.15$, and $\theta_{\rm imp} \simeq 0.48$~K. This amount of impurity matches the one found from the susceptibility fit. Further, after subtraction of the impurity contribution, the corrected magnetization stays at zero value up to $H_{\rm C1}$, as expected in the gapped state.


\subsection{Heat Capacity}
\begin{figure}[h]
\includegraphics[width= \linewidth]{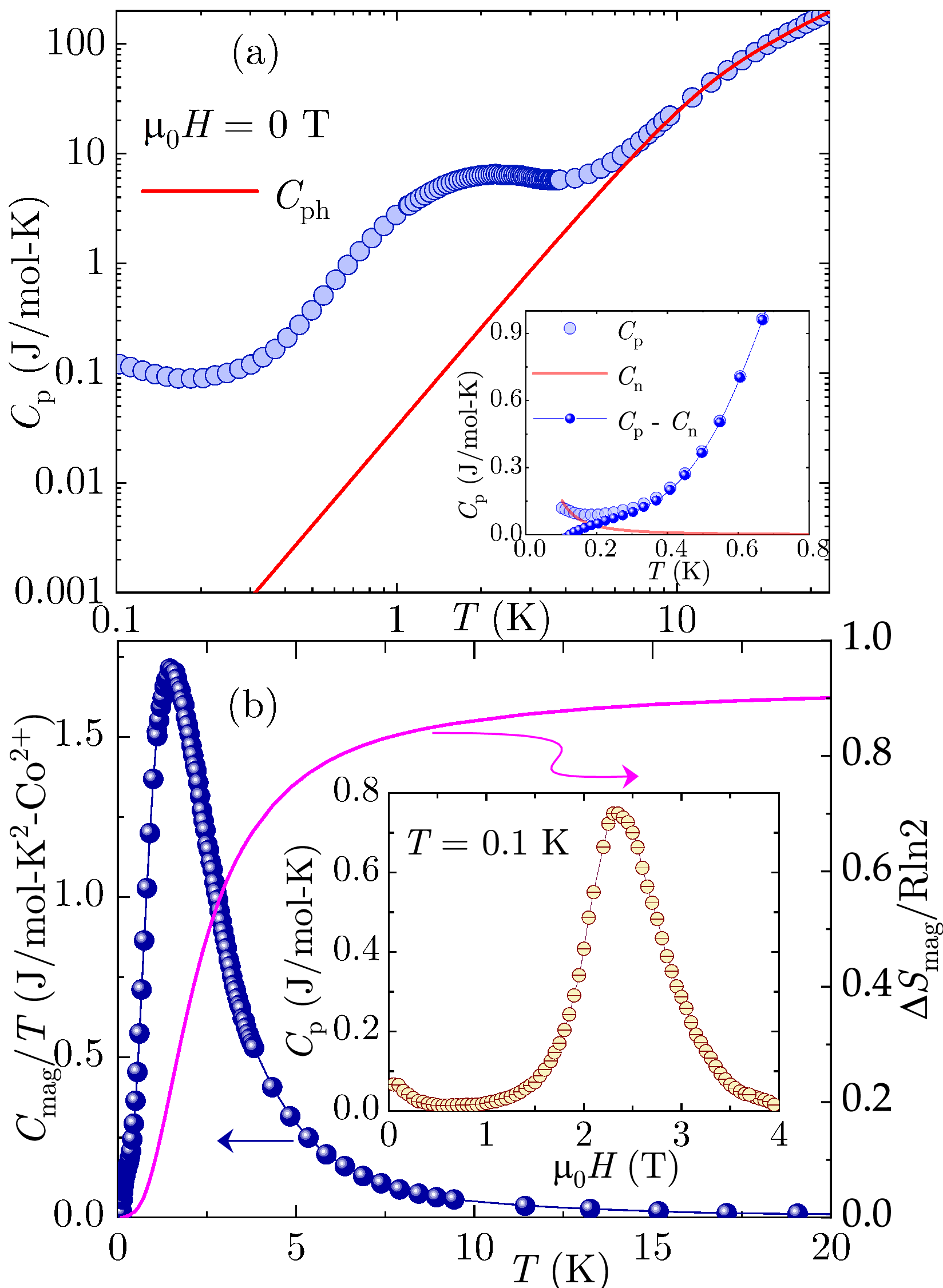}
\caption{\label{Fig4} (a) $C_{\rm p}$ vs $T$ measured in zero field along with the calculated phonon contribution $C_{\rm ph}$. Inset: Zero-field $C_{\rm p}(T)$ in the low-temperature regime with the solid line representing the nuclear contribution $C_{\rm n}$. (b) $C_{\rm mag}/T$ and $\Delta S_{\rm mag}$ vs $T$ on the left and right $y$-axes, respectively. Inset: $C_{\rm p}(H)$ measured at $T=0.1$~K.}
\end{figure}
Temperature-dependent heat capacity [$C_{\rm p}(T)$] measured in zero field and down to 100~mK is shown in Fig.~\ref{Fig4}(a). It depicts a well-defined broad maximum at around $\sim 3$~K, a hallmark of short-range AFM order. No signature of magnetic LRO is observed down to the lowest temperature. At very low-$T$s, a slight upturn is observed, which we attribute to the nuclear contribution appearing due to a small hyperfine field in this temperature regime, produced by Co or other atoms present. We estimated the nuclear contribution by fitting the zero-field $C_{\rm p}(T)$ data below 0.2~K with $C_{\rm n}(T) = \alpha_{\rm Q}/T^{2}$ [inset of Fig.~\ref{Fig4}(a)], where the coefficient $\alpha_{\rm Q}$ is related to the nuclear-level splitting, both quadrupolar and Zeeman. The fit yields $\alpha_{\rm Q} \simeq 1.6\times 10^{-3}$J\,K/mol, which aligns well with the values reported for other Co-based systems~\cite{Lal014429}. The data after subtraction of the nuclear contribution is also presented in the inset of Fig.~\ref{Fig4}(a) where $C_{\rm p}$ decreases towards zero on cooling, thus reflecting the singlet ground state.

In general, for a magnetic insulator, $C_{\rm p}(T)$ in zero field has two major contributions: phononic $C_{\rm ph}(T)$ and magnetic $C_{\rm mag}(T)$ parts. Above 10~K, $C_{\rm p}(T)$ is dominated by $C_{\rm ph}$, while at low temperatures, the dominant contribution is due to $C_{\rm mag}$. To estimate $C_{\rm ph}(T)$, experimental heat capacity data above 10~K were fitted empirically using a polynomial of the form~\cite{Nath054409}
\begin{equation}
C_{\rm ph}(T)= aT^3+bT^5+cT^7+dT^9.
\label{phonon}
\end{equation}
From the fit [Fig.~\ref{Fig4}(a)], the obtained coefficients are $a\simeq3.26\times10^{-2}$\,J\,mol$^{-1}$\,K$^{-4}$, $b\simeq 1.06\times10^{-4}$\,J\,mol$^{-1}$\,K$^{-6}$, $c\simeq2.04\times10^{-7}$\,J\,mol$^{-1}$\,K$^{-8}$, and $d\simeq-2.2\times10^{-10}$\,J\,mol$^{-1}$\,K$^{-10}$. $C_{\rm mag}$ was obtained by subtracting $C_{\rm ph}$ and $C_{\rm n}$ from the total $C_{\rm p}$. Furthermore, $C_{\rm mag}(T)$ was used to estimate the magnetic entropy [$\Delta S_{\rm mag}(T)$] by integrating $C_{\rm mag}(T)/T$ in the measured $T$-range. As shown in the right $y$-axis in Fig.~\ref{Fig4}(b), $\Delta S_{\rm mag}(T)$ $[= \int_{\rm 0.1\,K}^{T}\frac{C_{\rm mag}(T')}{T'}dT'$] reaches a well-defined plateau close to 90\% of $R\ln2$, expected for a pseudospin-$\frac12$ spin system. This again indicates that the low-temperature properties of Co-MOF are governed by the $J_{\rm eff}=1/2$ Kramers doublet~\cite{Lal014429}.

\begin{figure}[h]
\includegraphics[width= \linewidth]{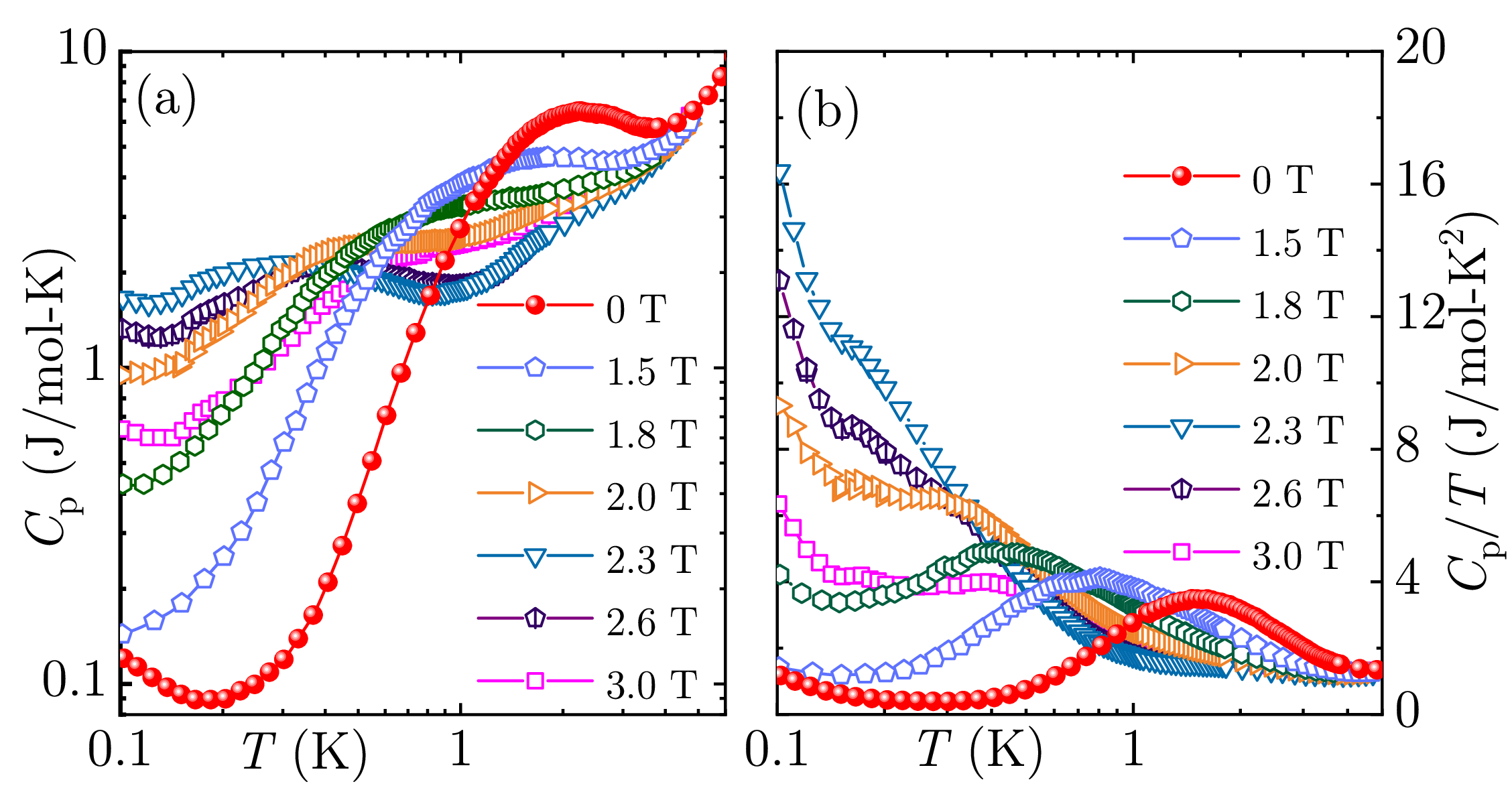}
\caption{\label{Fig5} (a) $C_{\rm p}$ vs $T$ measured at low temperatures in different applied fields. (b) $C_{\rm p}/T$ vs $T$ in the same temperature range.}
\end{figure}
In order to detect possible field-induced phases between $H_{\rm C1}$ and $H_{\rm C2}$, we also measured $C_{\rm p}(T)$ in different fields at low temperatures [see Fig.~\ref{Fig5}(a)]. With the application of field, the broad maximum shifts towards low temperatures, as expected for an AFM short-range-order. No feature/peak of field-induced magnetic LRO reminiscent of BEC of triplons is observed down to 100~mK for $H > H_{\rm C1}$. In order to visualize the absence of magnetic LRO, in Fig.~\ref{Fig5}(b), we plotted $C_{\rm p}/T$ vs $T$ measured in various applied fields that display no clear anomaly. The absence of a field-induced magnetic LRO suggests almost non-interacting dimers.

\subsection{Discussion}
\begin{figure}[h]
	\includegraphics[width= \linewidth]{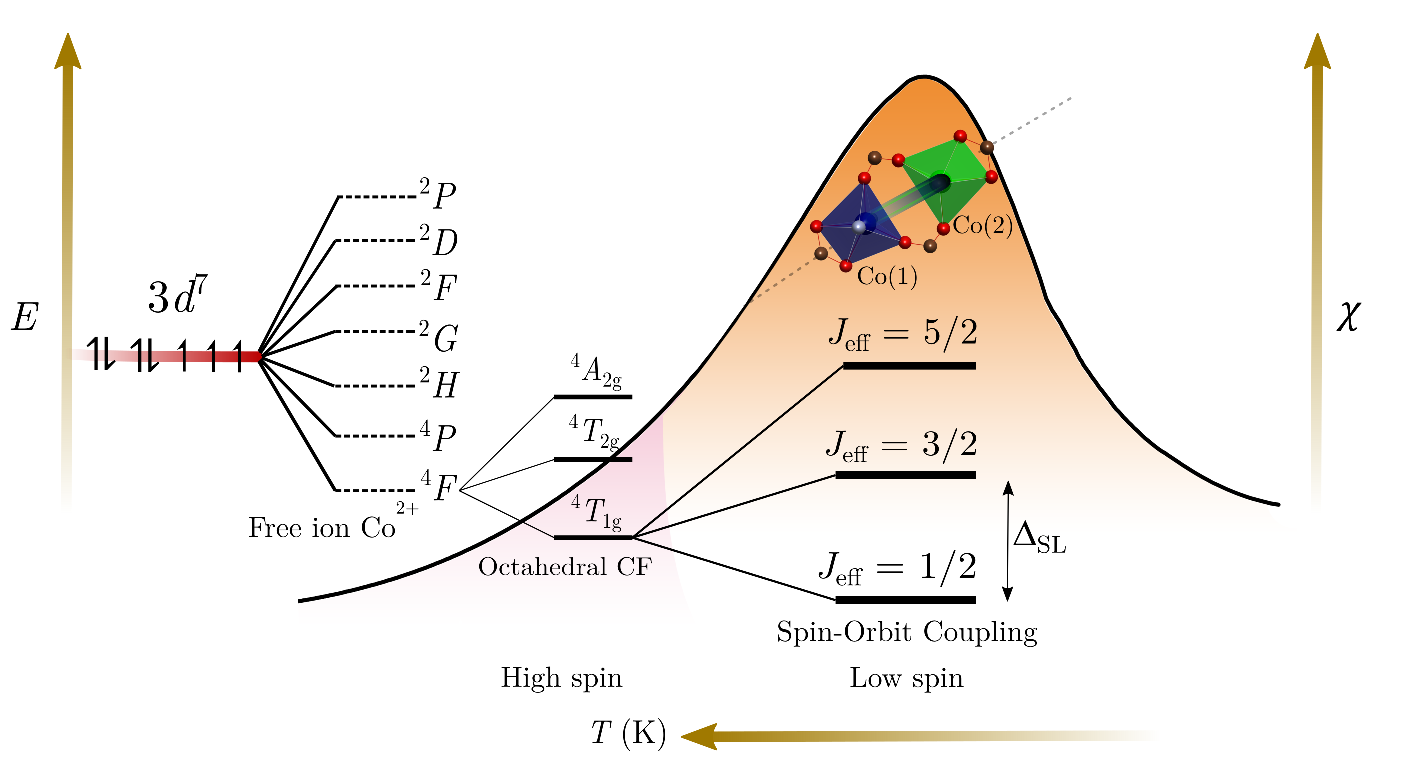}
	\caption{\label{Fig6} The tentative energy diagram that depicts the splitting of the ${^4}F$ state of Co$^{2+}$ under an octahedral crystal field and of the ${^4}T_{\rm 1g}$ state under SOC. $\chi(T)$ ($\mu_{0} H = 0.01$~T) is plotted in the right $y$-axis in order to visualize the temperature regimes where the high ($S=3/2$) and low-spin ($J_{\rm eff} = 1/2$) states are effective. The maximum in $\chi(T)$ is arising due to the spin dimers.}
\end{figure}
According to Hund's rule, the ground state of a free Co$^{2+}$ ion (3$d^{7}$) is ${^4}F$ with the orbital angular momentum $L=3$ and total spin $S=3/2$. 
This multiplet is split by the crystal field as well as SOC. In Fig.~\ref{Fig6}, we show the typical energy-level diagram of a Co$^{2+}$ ion in the crystal environment. Cubic crystal field (CF) splits the ${^4}F$ multiplet into three multiplets: ${^4}A_{\rm 2g}$, ${^4}T_{\rm 2g}$, and ${^4}T_{\rm 1g}$. For the octahedral geometry, the ${^4}T_{\rm 1g}$ multiplet is expected to be the ground state. The energy level (${^4}T_{\rm 1g}$) is further influenced by the SOC ($\hatH_{\rm SOC}=\lambda L \cdot S$), which splits this level into a doublet, quartet, and sextet (six Kramers doublets). The doublet is separated from the quartet by an energy gap of $\Delta_{\rm SL} = \frac{3}{2}\lambda$. At low temperatures, the corresponding degeneracy of the doublets is further lifted and the lowest Kramers doublet becomes the ground state. Thus, when the temperature is much lower than the spin-orbit coupling $\lambda$ (i.e. $T\ll |\lambda|/k_{\rm B}$), one expects the magnetism to be determined by the lowest Kramers doublet with the effective spin $J_{\rm eff}=1/2$~\cite{Abragam173,WellmL100420}. Indeed, all our thermodynamic measurements confirm the $J_{\rm eff} = 1/2$ ground state at low temperatures for Co-MOF. 

\begin{figure}[h]
	\includegraphics[width= \linewidth]{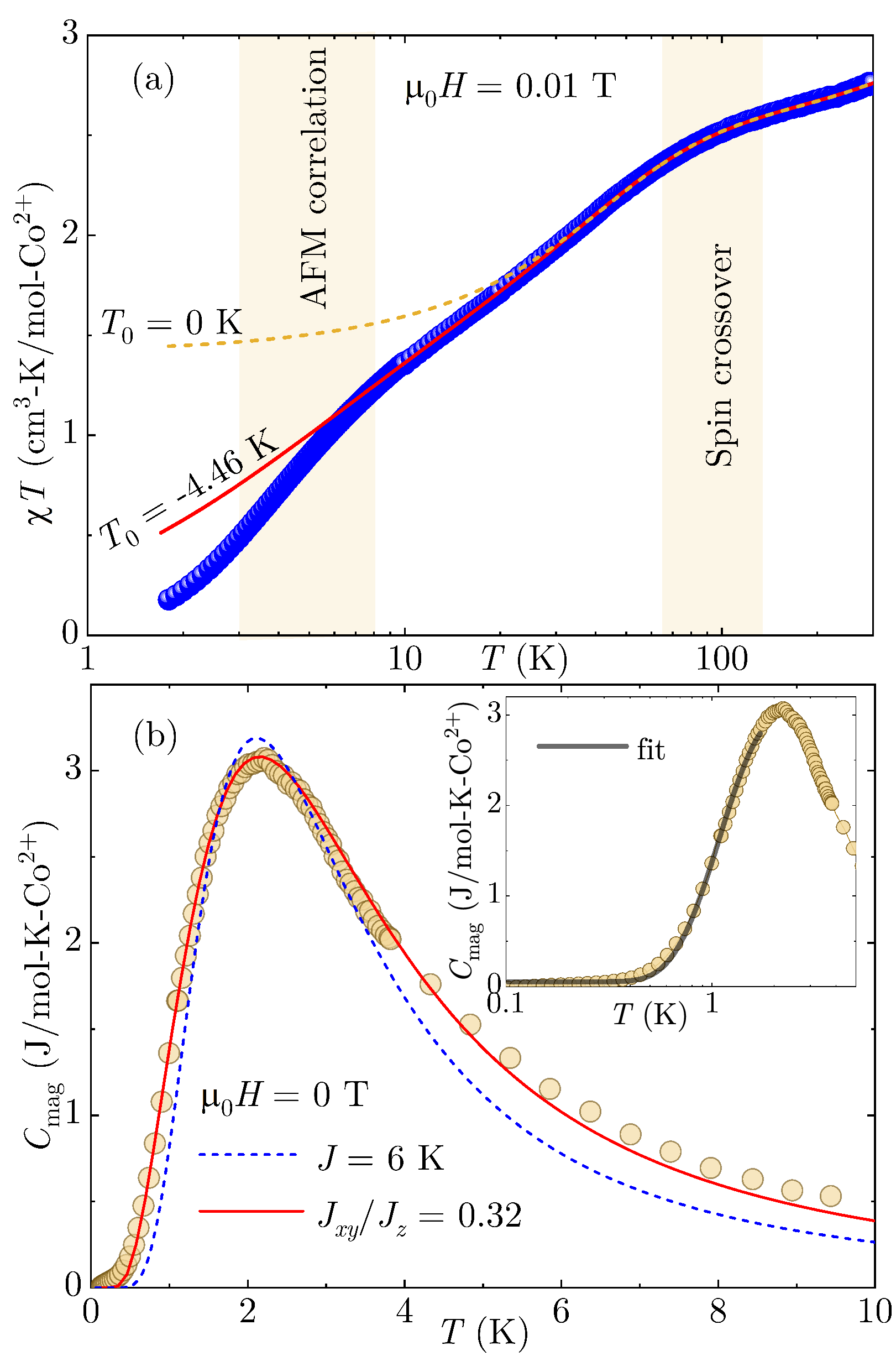}
	\caption{\label{Fig7} (a) $\chi T$ vs $T$ is plotted for $\mu_{0}H = 0.01$~T. The solid and dashed lines are fits using Eq.~\eqref{eq5} for $T_0 =0$~K and $T_0=-4.46$~K, respectively. (b) $C_{\rm mag}$ vs $T$ measured down to 0.1~K in zero field. The dashed and solid lines represent the simulated curves for an isolated spin-$1/2$ dimer with isotropic ($J = 6$~K) and anisotropic interactions ($J_{z} = 11$~K and $J_{xy} = 3.52$~K), respectively. Inset: Zero-field $C_{\rm mag}(T)$ in the low-$T$ regime; the solid line is an exponential fit with $\Delta_{0}^{\rm C}/k_{\rm B}\simeq 4.3$~K.}
\end{figure}
Upon revisiting the $\chi T$ vs $T$ data [Fig.~\ref{Fig7}(a)], we observed distinct slope changes that indicate changes in the effective magnetic moment of Co$^{2+}$. The temperature dependence can be fitted by a model of 
two lowest doublets separated by the gap $\Delta_{\rm SL}/k_{\rm B}$ as~\cite{Bartolom10382}
\begin{align}
\chi = \chi_0 +\frac{N}{k_{B}(T-T_{0})} \left[ C_{0} + C_{1}\left( \frac{2k_{\rm B}T}{\Delta_{\rm SL}} \right)\tanh\left( \frac{\Delta_{\rm SL} }{2k_{\rm B}T} \right) \right. \nonumber \\
\left.+ ~C_{2}\tanh\left( \frac{\Delta_{\rm SL} }{2k_{\rm B}T} \right) \right].
\label{eq5}
\end{align}
The value of $N$ represents the number of paramagnetic entities, the coefficients $C_0$, $C_1$, and $C_2$ depend on the electronic wave function of the two Kramers doublets involved, and $T_0$ essentially represents the energy scale of the exchange interactions. The fit above 10~K as shown in Fig.~\ref{Fig7}(a) returns $\chi_0\simeq 8.746\times10^{-4}$~cm$^3$/mol, $C_0/k_{\rm B} \simeq 0.972$~cm$^3$\,K/mol, $C_1/k_{\rm B}\simeq 0.668$~cm$^3$\,K/mol, $C_2/k_{\rm B}\simeq 0.249$~cm$^3$\,K/mol, and $\Delta_{\rm SL}/k_{\rm B}\simeq 160$~K with $T_0\simeq -4.4~$K. The large value of $\Delta_{\rm SL}$ implies that the system should behave as $J_{\rm eff} = 1/2$ at low temperatures ($T < \Delta_{\rm SL}/k_{\rm B}$). Indeed, our analysis below 30~K confirms this scenario. The negative value of $T_0$ suggests AFM correlations at low temperatures. In the low-$T$ regime when $T \ll \Delta_{\rm SL}/k_{\rm B}$, the second term within the parenthesis in Eq.~\eqref{eq5} is almost negligible and provides an effective Curie constant of $C_{\rm eff} = C_0/k_{\rm B}+C_2/k_{\rm B} \simeq 1.22$~cm$^3$-K/mol. This value corresponds to $\mu_{\rm eff} \simeq 3.12~\mu_{\rm B}$ that matches closely with $\mu_{\rm eff}^{\rm LT}$ obtained from the CW fit.

Structurally, Co-MOF presents an ideal case of isolated dimers. Though the isolated dimer model with the isotropic (Heisenberg) intradimer interaction fits the $\chi(T)$ data reasonably well, it fails to reproduce the shape of the magnetization isotherm and $C_{\rm mag}(T)$. A more narrow maximum is expected in the heat capacity of a Heisenberg spin dimer. Even if we choose $J = 6$~K instead of $J = 7.6$~K from the
$\chi(T)$ fit in order to reproduce the position of the maximum,
its height should be reduced by 25\%, which is inconsistent
with the almost full magnetic entropy recovered in our measurements.
Despite all these adjustments, the peak remains more narrow than in the
experiment, thus suggesting that Co-MOF can not be treated as Heisenberg
spin dimer.

The heteroleptic, distorted CoO$_4$N$_2$ octahedron is elongated along the $z$-direction as compared to the $xy$-plane. This implies that a trigonal crystal-field splitting is likely to develop. It leads to a Hamiltonian of the form $\hatH_{\rm trig} = \delta (3L_{\rm z}^2 -2)$, where $\delta$ denotes the strength of CFS~\cite{Liu047201}. It is theoretically predicted that under the influence of trigonal distortion, isotropic Heisenberg exchange gives way to an anisotropic one described by the XXZ model~\cite{Liu054420}. Therefore, in Fig.~\ref{Fig7}(b) we simulated the $C_{\rm mag}(T)$ using such a model,
\begin{equation}
\mathcal \hatH=J_{xy}(S_i^xS_j^x+S_i^yS_j^y)+J_{z}S_i^zS_j^z,
\label{eq:ham}
\end{equation}
with the exchange couplings $J_{xy}\simeq 3.5$~K and $J_{z}\simeq 11$~K. The anisotropy of $J_{xy}/J_{z}\simeq 0.32$ controls the width and height of the maximum, so it can be determined with relative accuracy even from powder data. With the XXZ model, we are also able to replicate $\chi(T)$ and $M$ vs $H$ curves using the same exchange parameters assuming isotropic $g=4.15$ (see Fig.~\ref{Fig3}) and the same impurity contribution as in Sec.~\ref{sec:magnetization}.
The average exchange coupling $J_{\rm avg}/k_{\rm B}= \frac{2J_{xy}+J_{z}}{3k_{\rm B}} \simeq 6$~K with $g = 4.15$ corresponds to $H_{\rm C1} \simeq 2$~T. Similar physics of isolated spin dimers with anisotropic interactions is also realized in $4f$ systems, such as BiYbGeO$_5$ and NaLu$_{0.9}$Yb$_{0.1}$Se$_2$~\cite{Mohanty134408,Cairns024404}.

We further quantified the spin gap by fitting $C_{\rm mag}(T)$ by the low-$T$ approximation of the spin-$1/2$ dimer model: $C_{\rm mag} \propto \left(\Delta_{0}^{\rm C}/k_{\rm B}T\right)^2 e^{-\Delta_{0}^{\rm C}/k_{\rm B}T}$~\cite{Freitas184426}. The fit below 2~K [inset of Fig.~\ref{Fig7}(b)] returns $\Delta_{0}^{\rm C}/k_{\rm B} = 4.3(1)$~K and this value seems to be reasonable when compared with $J_{\rm avg} \simeq 6$~K of the anisotropic XXZ model.

The scenario of isolated spin dimers are also further confirmed from the magnetization isotherm and $C_{\rm p}$ vs $H$ measurements. In a magnetization isotherm, the spacing between $H_{\rm C1}$ and $H_{\rm C2}$ is normally controlled by the magnitude of the inter-dimer coupling $J'/k_{\rm B}$, i.e., increasing the value of $J'/k_{\rm B}$ widens the spacing between them. In such a scenario, the derivative of $M$ ($dM/dH$) vs. $H$ should exhibit two peaks corresponding to $H_{\rm C1}$ and $H_{\rm C2}$~\cite{Lancaster207201}. As presented in Fig.~\ref{Fig3}(b), our $dM/dH$ vs $H$ plot reveals only one peak at around $2.2$~T, reflecting the fact that the compound is an isolated dimer system with negligible inter-dimer coupling~\cite{Tsirlin104436}. $C_{\rm p}$ vs $H$ measured at the lowest temperature of $T = 0.1$~K is shown in the inset of Fig.~\ref{Fig4}(b). $C_{\rm p}$ remains almost zero in low fields due to the singlet ground state. As the increasing field approaches $H_{\rm C1}$, $C_{\rm p}$ starts increasing, shows a peak at around 2.2~T, and then decreases further to zero in higher fields~\cite{Cairns024404}.
Usually, in a system of coupled dimers with the spin gap, $C_{\rm p}$ vs $H$ shows double peaks corresponding to $H_{\rm C1}$ and $H_{\rm C2}$~\cite{Freitas184426,Kohama037203}. However, for an isolated dimer with negligible interdimer coupling, one may expect only one peak. In our case, we observed a single broad peak in $C_{\rm p}(H)$ at $T=0.1$~K that clearly indicates the proximity of our system to the isolated-dimer model where both $H_{\rm C1}$ and $H_{\rm C2}$ coincide when $T\rightarrow{0}$.

\subsection{Conclusion}
Co-MOF features $J_{\rm eff} = 1/2$ Co$^{2+}$ dimers arranged in an orthogonal fashion, similar to the famous Shastry–Sutherland lattice. In zero field, a quantum disordered ground state with the spin gap and no magnetic LRO is observed. The absence of magnetic LRO in applied fields down to 100~mK excludes BEC of triplons and underpins the absence of any significant interdimer coupling. We observed that the ground-state Kramers doublet is influenced by both trigonal CFS and SOC, causing the spin system to deviate from the Heisenberg dimer limit. Indeed, the low-temperature properties are well described by an isolated spin-1/2 anisotropic XXZ dimer model ($J_{xy} \neq J_{z}$). No magnetization plateau is observed despite the presence of orthogonal dimers. However, this compound can serve as a model system to study magnetization plateaus if one can introduce a finite inter-dimer coupling by an appropriate choice of organic ligands.

\acknowledgements
We would like to acknowledge SERB, India for financial support bearing sanction Grant No.~CRG/2022/000997. SJS is supported by the Prime Minister’s Research Fellowship (PMRF) scheme, Government of India. Computations for this work were done using resources of the Leipzig University Computing Center.


%

\end{document}